\documentclass[twocolumn,           
               showpacs,            
               preprintnumbers,     
               aps,                 
               prd,                 
               superscriptaddress,      
               nofootinbib,         
               tightenlines,        
               floats,floatfix      
               ]{revtex4}

\usepackage{graphicx}
\usepackage{color}
\usepackage{latexsym}
\usepackage{amsmath,amssymb}        
\usepackage[draft=false]{hyperref}

\begin{document}


\title{Spherical Boson Stars as Black Hole mimickers}

\author{F. S. Guzm\'an}
\affiliation{Instituto de F\'{\i}sica y Matem\'{a}ticas,
        Universidad Michoacana de San Nicol\'as de Hidalgo. Edificio C-3,
        Cd. Universitaria,
        C. P. 58040 Morelia, Michoac\'{a}n, M\'exico.}

\author{J. M. Rueda-Becerril}
\affiliation{Instituto de F\'{\i}sica y Matem\'{a}ticas,
        Universidad Michoacana de San Nicol\'as de Hidalgo. Edificio C-3,
        Cd. Universitaria,
        C. P. 58040 Morelia, Michoac\'{a}n, M\'exico.}
\affiliation{Facultad de Ciencias,
                Universidad Aut\'onoma del Estado de M\'exico,
                Instituto Literario 100, Toluca, Edo. Mex.,
                50000 M\'exico.}


\date{\today}


\begin{abstract}
We present spherically symmetric boson stars as black hole mimickers based on the power spectrum of a simple accretion disk model. The free parameters of the boson star are the mass of the boson and the fourth order self-interaction coefficient in the scalar field potential. We show that even if the mass of the boson is the only free parameter it is possible to find a configuration that mimics the power spectrum of the disk due to a black hole of the same mass. We also show that for each value of the self-interaction a single boson star configuration can mimic a black hole at very different astrophysical scales in terms of the mass of the object and the accretion rate. In order to show that it is possible to distinguish one of our mimickers from a black hole we also study the deflection of light.
\end{abstract}




\maketitle



\section{Introduction}
\label{sec:introduction}

Due to a number of observations related to high energy events, at present time one important problem in relativistic astrophysics is that of the nature of black hole candidates (BHCs). It is usually assumed that black hole solutions are the only models of BHCs. This together with the construction of various non-vacuum solutions in general relativity for various types of matter and configurations, the question of whether BHCs are black hole solutions or a different solution has been of certain interest. Among such alternative solutions are wormholes \cite{Harko2009,Lemos2008}, gavastars \cite{Visser2004,Rezzolla2007,Lemos2008,Harko2009c}, brane world solutions \cite{Harko2008} and boson stars \cite{diego,diego-acc,Guzman2006}. The reason is that astrophysical implications of these solutions cannot be distinguished from those due to black hole solutions at distances far from the event horizon region and thus the existence of the horizon itself has become the object of study of BHCs \cite{narayan}.

Various properties of black hole mimickers are to be studied, being in the first place the  stability. For instance, stability of the whole set of wormhole solutions usually thought of as mimickers is not clear, instead the instability of wormhole solutions has been shown for basic wormhole solutions \cite{Hayward,GGS1,GGS2,GGS3}, destroying previous hope on the possibility that these are stable as shown in \cite{Armendariz} for particular types of perturbations.On the other hand, the stability of gravastars has been explored and has been found that there are stability regions \cite{Visser2004,Carter2005,Rezzolla2008}. Also on favor of boson star solutions it can be said that the stability of the solutions has actually been exhaustively studied, for instance using perturbative methods \cite{Gleiser,scott}, catastrophe theory \cite{Catastro} and full non-linear numerical relativity, both in spherical symmetry  \cite{SeidelSuen1990,Balakrishna1998,scott} and full 3D \cite{Guzman2004}, and the stable branches of solutions are well known.  

In fact, the study of boson stars has been pushed forward up to the binary boson star collision \cite{Lehner2007} and the non-linear evolution of perturbed boson stars \cite{ruxandra}, in both cases  considering the system as a source of gravitational waves. The study of gravitational wave signatures has also been studied in order to tell between a black hole and a gravastar \cite{Rezzolla2007}. Instead, in the case of wormholes e.g. simple solutions (supported by a phantom scalar field) would not stand the perturbative analysis and there is no hope for a study of a binary system,  because as shown in \cite{GGS2} the life time of these solutions is rather short and they should either collapse and form black holes or explode before they could merge, although in \cite{GGS3} it was shown that  charged wormholes can have a longer life-time if the charge parameter is adequately chosen. About a potential full non-linear study of gravastars, the challenges are related to the evolution of the fluid and numerical methods should be able to handle very well located distributions of matter, nevertheless some alternative gravastar-like solutions have been proposed to ameliorate the infinitesimally thin shell problem by introducing anisotropies in the fluid \cite{Cattoen2005}. 
At the end of the day, if black hole mimickers are to be compared from all the angles with black holes, mimickers should also be expected to do what black holes can do: they collide and generate gravitational radiation with a given fingerprint.

The black hole mimicker models mentioned above are usually compared using the study of the accretion of geometrically thin optically thick disk models, since they are based on the study of time-like geodesics on a fixed background space-time. In this paper we present a set of boson star solutions that can act as black hole mimickers based on the same accretion disk model as done for the other mimickers. We show that they appear to be independent of the astrophysical parameters related to the mass of the BHC and the accretion rate. With these results we find that the power spectrum obtained out of this type of accretion disk model does not suffice to distinguish a boson star from a black hole if the boson star is chosen appropriately.
This paper is a generalization of \cite{Guzman2006}, in the following sense: 1) we show that a boson star mimicker for one BHC mass is also the same mimicker for other BHC mass by only changing the values of the boson mass, 2) we track the mimickers for various values of the self-interaction parameter and 3) we present the deflection of light as a method to eventually distinguish a black hole from a boson star.

In order to provide potential predictions that help distinguishing our boson stars from a black hole, we study the light deflection by our boson stars and by the mimicked black hole, show that in the boson star case there is no photon sphere as expected and show the scale at which high resolution lensing would distinguish between the boson star and the black hole.

The paper is organized as follows: 
in section \ref{sec:bosonstars}
we present the construction of boson star solutions. 
In section III we describe the disk model and the calculation of the power spectrum.
In section IV we present the boson stars able to mimic black holes whereas in section V we present a prediction to falsify our model based on the deflection of light by boson stars.
Finally in section VI we draw some conclusions.


\section{Spherical Boson Stars}
\label{sec:bosonstars}

Boson stars (BSs)  are solutions to Einstein's field equations minimally coupled to a complex scalar field \cite{Kaup1968,RuffiniBonazzola1969}. The scalar field is endowed with a potential of self-interaction that guarantees a global U(1) symmetry. This global phase invariance implies 
the existence of a conserved charge corresponding to the boson
particle number of equilibrium configurations.
The Lagrangian density describing a minimally coupled complex scalar 
field to general relativity reads:

\begin{equation}
{\cal L} = -\frac{R}{2 \kappa_0} + g^{\mu \nu}\partial_{\mu} \phi^{*}
\partial_{\nu}\phi + V(|\phi|^2),
\label{eq:lagrangian}
\end{equation}

\noindent where $\kappa_0 = 8 \pi$ in units where $c=G=1$, $\phi$ is the
scalar field, the star stands for complex conjugate and $V$ is the potential of 
the scalar field. Notice that this Lagrangian density
is invariant under the global $U(1)$ group, and the associated
conserved charge is called the number  of particles (defined below). When
the action is varied with respect to the metric
Einstein's equations arise $G_{\mu\nu} = \kappa_0 T_{\mu\nu}$, where the
stress-energy tensor reads

\begin{equation}
T_{\mu \nu} = \frac{1}{2}[\partial_{\mu} \phi^{*} \partial_{\nu}\phi +
\partial_{\mu} \phi \partial_{\nu}\phi^{*}] -\frac{1}{2}g_{\mu \nu}
[\phi^{*,\alpha} \phi_{,\alpha} + V(|\phi|^2)].
\label{eq:set}
\end{equation}

In the case of boson stars as black hole mimickers, it suffices to consider the potential of the type $V(|\phi|) = m^2|\phi|^2 + \lambda |\phi|^4$ where $m$ is associated to the mass of the boson and $\lambda$ to the self-interaction of the boson system. Finally, Bianchi identities reduce to the Klein-Gordon equation

\begin{equation}
\left(
\Box - \frac{dV}{d|\phi|^2}
\right)\phi = 0,
\end{equation}

\noindent where $\Box
\phi=\frac{1}{\sqrt{-g}}\partial_{\mu}[\sqrt{-g}
g^{\mu\nu}\partial_{\nu}\phi]$.

Boson stars are spherically symmetric solutions of the equations above when a harmonic
time-dependence for the scalar field is assumed $\phi(r,t) = \phi_0(r) e^{-i \omega t}$,
where $r$ is the radial coordinate and $t$ the coordinate time. This condition implies that
the stress-energy tensor becomes time-independent, which in turn implies that the geometry
of the space-time is also time-independent.


Boson star solutions are constructed considering the time-independent spherically symmetric line element

\begin{equation}
ds^2=-\alpha(r)^2dt^2 + a(r)^2dr^2 + r^2 d\Omega^2,
\label{eq:metric}
\end{equation}

\noindent with $\alpha(r)$ the lapse function and $a(r)$ the
radial metric function.
Under these specifications, the Einstein-Klein-Gordon equations become:

\begin{eqnarray}
\frac{\partial_r a}{a} &=& \frac{1-a^2}{2r} +\nonumber\\
        &&\frac{1}{4}\kappa_0 r
        \left[\omega^2 \phi_{0}^{2}\frac{a^2}{\alpha^2}
        +(\partial_r \phi_0)^{2} +
        a^2 \phi_{0}^{2} (m^2 + \lambda \phi_{0}^{2})
        \right],\nonumber\\
\frac{\partial_r \alpha}{\alpha} &=&
        \frac{a^2-1}{r} +
        \frac{\partial_r a}{a} -
        \frac{1}{2}\kappa_0 r a^2\phi_{0}^{2}(m^2 
        +\lambda \phi_{0}^{2}),\nonumber\\
\partial_{rr}\phi_0  &+& \partial_r \phi_0  \left( \frac{2}{r} +
        \frac{\partial_r \alpha}{\alpha} - \frac{\partial_r a}{a}\right)
        + \omega^2 \phi_0 \frac{a^2}{\alpha^2} \nonumber\\
        &-& a^2 (m^2 + 2\lambda \phi_{0}^{2}) \phi_0
        =0.\label{sphericalekgc-sc}
\end{eqnarray}

\noindent The system (\ref{sphericalekgc-sc}) is a set of
coupled ordinary differential equations to be solved under the
conditions $a(0)=1$ that provides spatial flatness, $\phi_0(0)$
finite and $\partial_r \phi_0(0)=0$ as a condition that guarantees
the regularity of the operators at the origin;
also demand $\phi_0(\infty)=0$ in order to ensure asymptotic
flatness at infinity. 
The problem turns into an eigenvalue problem for $\omega$, and for a
given value of the central field $\phi_0$ there is a unique $\omega$ with which the
boundary conditions are satisfied. In order to get rid of the constants in the equations we
re-scale variables and constants by
$\tilde{\phi}_0 = \sqrt{\frac{\kappa_0}{2}} \phi_0$,
$\tilde{r} = m r$,
$\tilde{t} = \omega t$,
$\tilde{\alpha} = \frac{m}{\omega}\alpha$ and
$\Lambda = \frac{2\lambda}{\kappa_0 m^2}$.
In terms of these new constants, and after
removing the tildes from the variables the system of equations becomes

\begin{eqnarray}
\frac{\partial_r a}{a} &=& \frac{1-a^2}{2r} +\nonumber\\
        &&\frac{1}{2} r
        \left[\phi_{0}^{2}\frac{a^2}{\alpha^2}
        +(\partial_r \phi_0)^{2} +
        a^2 (\phi_{0}^{2}
        + \Lambda \phi_{0}^{4})
        \right],\nonumber\\
\frac{\partial_r \alpha}{\alpha} &=&
        \frac{a^2-1}{r} +
        \frac{\partial_r a}{a} -
        r a^2\phi_{0}^{2}(1 
        +\Lambda\phi_{0}^{2}),\nonumber\\
\partial_{rr}\phi_0  &+& \partial_r \phi_0
        \left(
                \frac{2}{r} +
                \frac{\partial_r \alpha}{\alpha} - \frac{\partial_r a}{a}
        \right)
        + \phi_0 \frac{a^2}{\alpha^2} \nonumber\\
        &-& a^2 (1 + 2\Lambda \phi_{0}^{2}) \phi_0
        =0.
\label{sphericalekgc-sc-rescaled}
\end{eqnarray}

\noindent Notice that $\omega$ now turns into
the central value of the lapse $\alpha$ due to the rescaling. This is the
system we solve using finite differences with an ordinary
integrator (adaptive step-size fourth order Runge-Kutta algorithm in the present case) and a shooting routine
that bisects the value of $\omega$ (the central value of $\alpha$).
Also, it is worth noticing that the radial coordinate is scaled by $m$ and thus a
natural scale for our calculations will be given by $m=1$ which together with $\Lambda$ are the two parameters determining the properties of each configuration.


\begin{figure}[htp]
\includegraphics[width=8cm]{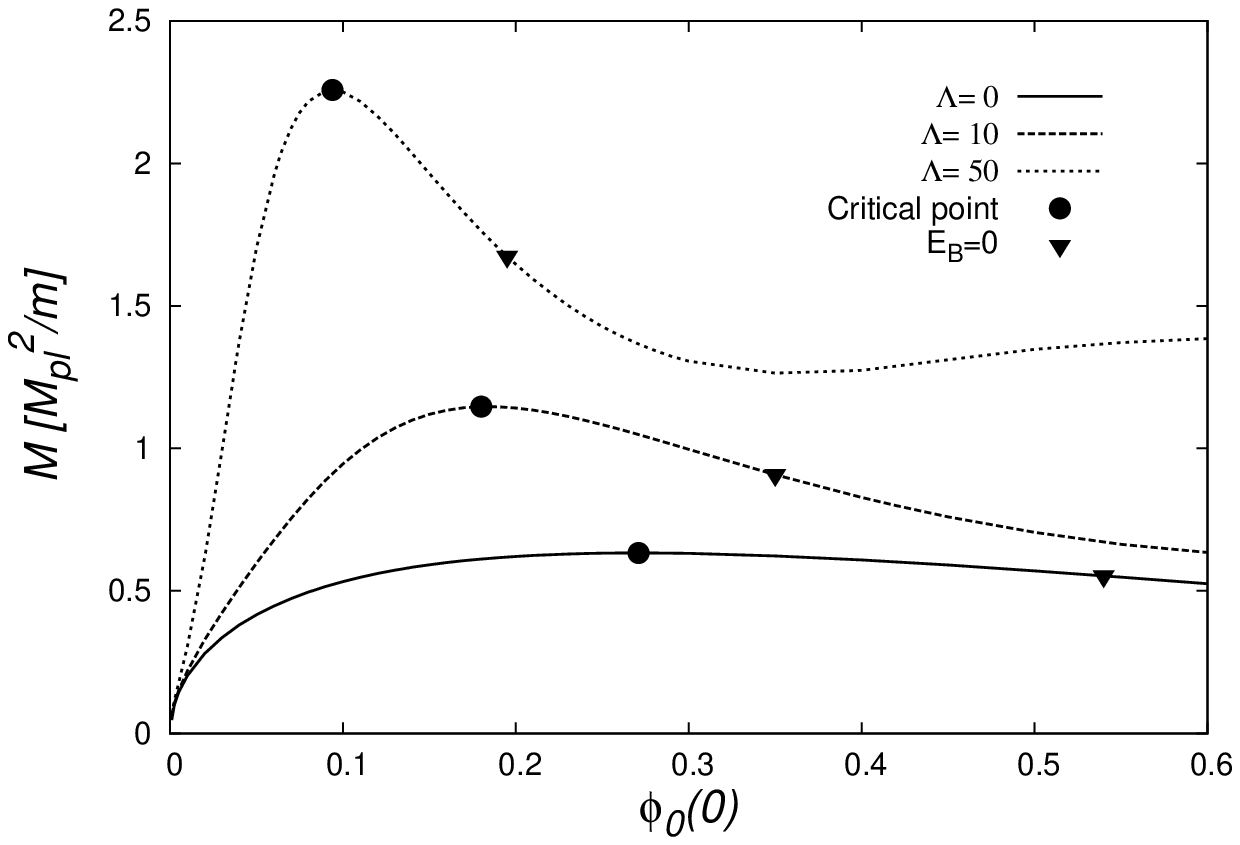}
\includegraphics[width=8cm]{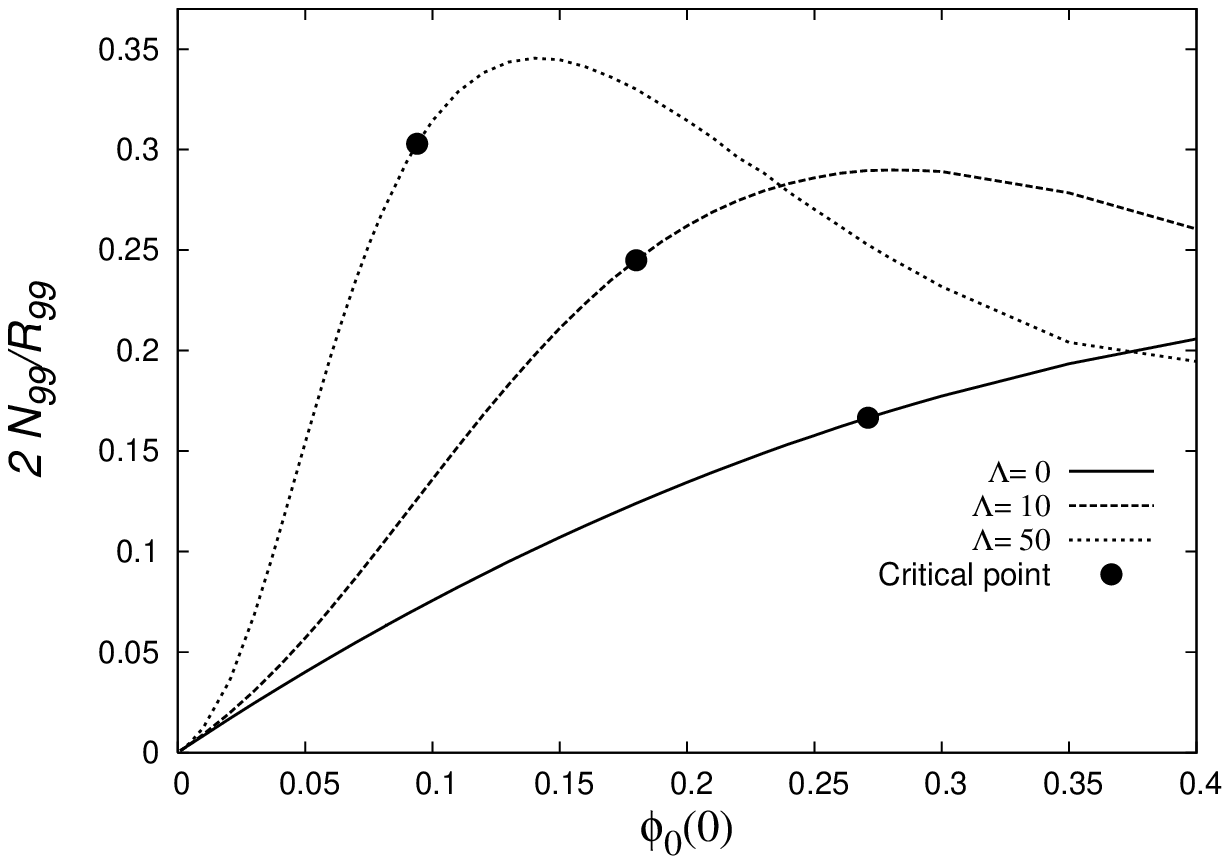}
\caption{\label{fig:equilibrium} (Top) Sequences of equilibrium
configurations for different values of $\Lambda$ are shown as function
of the central value of the scalar field $\phi_0(0)$;
each point in the curves corresponds to a solution of the
eigenvalue problem and represents a boson star.
The filled circles indicate the critical solution that
separates the stable from the unstable branch. Those
configurations to the left of the maxima represent stable
configurations. The inverted triangles
indicate the point at which the binding energy is zero. The mass of the configurations is used in units of the squared Planck mass over the mass of the boson, which implies a scale invariance of this plot under the change of $m$.
(Bottom) We show the compactness of the equilibrium configurations shown. We define compactness as the ratio $2N_{99}/R_{99}$, where $R_{99}$ is the radius at which 99 percent of the total number of particles $N_{99}$ is contained. In this plot we also show the critical point with a filled circle and indicate that the critical configurations are the most compact ones among the stable configurations for each value of $\Lambda$. Those configurations to the right of the critical solutions are even more compact however unstable and are not considered here. In our results below we will also use the critical solutions for reference as the most compact stars.}
\end{figure}

The solutions of (\ref{sphericalekgc-sc-rescaled}) define sequences of
equilibrium configurations like those shown in Fig.
\ref{fig:equilibrium}. Each point in the curves corresponds to a
boson star solution. In each of the curves two
important points for the different values of $\Lambda$ are indicated: i) the critical
point -marked with a filled circle- indicating the threshold between the
stable and unstable branches of each sequence, that is, configurations
to the left of this point are stable and those to the right are unstable
as found for boson stars in the full non-linear regime \cite{SeidelSuen1990,Balakrishna1998,Guzman2004}
and ii) the point at which the binding energy $E_B = M-Nm = 0$, indicated
with an inverted filled triangle, where $N=\int
j^0 d^3 x = \int
\frac{i}{2}\sqrt{-g}g^{\mu\nu}[\phi^{*}\partial_{\nu}\phi - \phi
\partial_{\nu}\phi^{*}]d^3 x$ is the number of particles; that is,
the conserved quantity due to the invariance under the global $U(1)$
group of the Lagrangian density (\ref{eq:lagrangian}). For a mass estimate we use the Misner-Sharp mass defined through the metric functions by $M(r) =
(1-1/a^2)r/2$, being $M$ the mass function evaluated at the outermost
point of the numerical domain; the configurations between the instability
threshold and the zero binding energy point have negative binding energy
($E_B<0$) and collapse into black holes, whereas those to the right
of the inverted triangles have positive binding energy and disperse away \cite{Guzman2004}. Those
configurations to the left of the threshold of instability, that is,
stable configurations, obviously posses negative binding energy, and are the expected ones to be studied as black hole mimickers. For reviews and more information on boson stars see 
\cite{Jetzer1992,topical-review}.


\section{Disk model and the power spectrum}
\label{sec:spectra}

Following a procedure similar to that described in \cite{diego-acc} to calculate the power spectrum from the disk model, we first study the geodesics. Given the line element $ds^2 = -\alpha(r)^2
dt^2 + a(r)^2dr^2 + r^2d\Omega^2$, the equation for time-like geodesics
followed by test particles on the equatorial plane reads:
               
\begin{equation}
\dot{r}^2 + \frac{1}{a^2}\left(1 + \frac{L^2}{r^2}\right) =
\frac{E^2}{\alpha^2 a^2},
\label{eq:geodesics}
\end{equation}

\noindent  where $L^2 = r^4\dot{\varphi}^2$ and $E^2 = -
\alpha^2 \dot{t}^2$ are the squared angular momentum and energy
at spatial infinity, which are the conserved quantities of the test
particle related to the independence on the azimuthal angle $\varphi$ and
$t$ of the space-time respectively; an overdot indicates derivative with
respect to the proper time of the test particle. The geodesics for a
Schwarzschild black hole are given by (\ref{eq:geodesics}) with the values
$\alpha^2=a^{-2}=(1-\frac{2M}{r})$. Since the equation
(\ref{sphericalekgc-sc-rescaled})  for $\alpha$ is linear, we rescale this function
so that at infinity $\alpha(r \rightarrow \infty) = 1/a(r \rightarrow
\infty)$, thus at infinity $\alpha^2 = a^{-2}$, which implies the
coefficient of $E^2$ in (\ref{eq:geodesics}) equals one for both boson stars and black holes.

The study of stable orbits of test particles requires the effective potential
$V_{eff}^{2} = \frac{1}{a^2}\left(1 + \frac{L^2}{r^2}\right)$ for a given
angular momentum of the test particle. In order to consider only circular orbits we rewrite (\ref{eq:geodesics}) for $\dot{r}=0$ and demand $dV/dr=0$ with
$V=\left[ \left( 1 + L^2/r^2 \right) - E^2/\alpha^2\right]/a^2$. This
condition implies

\begin{equation}
E = \frac{\alpha^2}{\sqrt{\alpha^2 - r\alpha\alpha^{\prime}}}, ~~~\&~~~
L = \sqrt{\frac{r^3\alpha\alpha^{\prime}}{\alpha^2 -
r\alpha\alpha^{\prime}}}
\label{eq:EyL}
\end{equation}

\noindent where $\alpha^2 - r\alpha\alpha^{\prime} > 0$ for all BSs.
Before proceeding to the calculation of the emission spectrum we need the
angular velocity of a test particle given by $\Omega = \sqrt{\frac{\alpha
\alpha^{\prime}}{r}}$, which is a regular function for all values of $r$ in the boson star case.

The accretion disk model is that of a geometrically thin, optically thick,
steady accretion disk. The power per unit area generated by such a disk
rotating around a central object is given by \cite{page-thorne,diego-acc} and used also in the study of disks around gravastars \cite{Harko2009c}:

\begin{equation}
D(r) = \frac{\dot{M}}{4\pi r}\frac{\alpha}{a}\left(-\frac{d\Omega}{dr}
\right)
\frac{1}{(E-\Omega L)^2} \int^{r}_{r_{i}}(E-\Omega L)\frac{dL}{dr}dr,
\label{eq:power}
\end{equation}

\noindent where $\dot{M}$ is the accretion mass rate and $r_{i}$ is the
inner edge of the disk. For
black holes this radius is assumed to be at the ISCO ($r=6M$) of the hole.
For BSs we choose $r_i=0$ based on two considerations: i) BSs allow circular orbits in the whole spatial domain and thus there is no kinematical restriction to choose  $r_{i} = 0$ and ii) it has been shown that the luminosity of the disk for BSs considering this inner radius never reaches the value of Eddington luminosity in the whole spatial domain even for high accretion rates and thus there are no radiation pressure effects imposing a restriction on the inner edge location \cite{diego-acc}. Considering the disk starting at $r_i=0$ does not affect the boson star structure as long as the disk is made of test particles;  the case where the interaction of the disk matter with the geometry of the space-time is taken into account -which is beyond the scope of this paper- may impose restrictions to the geometrical properties of the disk. Moreover, considering other values of $r_i$ for the BSs within the present model permits eventually to find also another configuration that would act as a mimicker. 
Now, assuming it is possible to define a
local temperature we use the Steffan-Boltzmann law so that $D(r) = \sigma
T^4$, where $\sigma=5.67 \times 10^{-5}~ erg~s^{-1}~cm^{-2}~K^{-4}$ is the
Steffan-Boltzmann constant. Now, considering the disk emits as a black
body, we use the dependence of $T$ on the radial coordinate
and therefore the luminosity $L(\nu)$ of the disk and the flux $F(\nu)$ can be calculated 
using the expression for the black body spectral distribution:

\begin{equation}
L(\nu) = 4\pi d^2 F(\nu) =
\frac{16 \pi h}{c^2} \cos (\vartheta) \nu^3 \int^{r_f}_{r_i}
\frac{rdr}{e^{h\nu/kT} - 1},
\label{eq:luminosity}
\end{equation}

\noindent where $d$ is the distance to the source, $r_i$ and $r_f$
indicate the location of the inner and outer edges of the disk, $h=6.6256
\times 10^{-27}~ erg~s$ is the Planck constant, $k = 1.3805 \times
10^{-16}~ erg~K^{-1}$ is the Boltzmann constant and $\vartheta$ is the
disk inclination. The algorithm to construct the emission spectrum for
such a model of accretion disk around a BS and a BH is as follows: \\

\noindent 1) Define the space-time functions $a$ and $\alpha$ by choosing
one of the equilibrium configurations in Fig.~\ref{fig:equilibrium} and
calculate $M$.\\
2) Define the metric of the equivalent BH through $\alpha_{BH} =
\sqrt{1-2M/r}$ and $a_{BH} = 1/\sqrt{1-2M/r}$.\\
3) Calculate the angular velocity, angular momentum and energy of a test
particle for both space-times $\Omega_{BS, BH}$, $L_{BS,BH}$,
$E_{BS,BH}$.\\
4) Use such quantities to calculate the power emitted in both cases
$D_{BS}(r)$ and $D_{BH}(r)$ defined in (\ref{eq:power}).\\
5) Calculate the temperature of the disk in both cases
$T_{BS}(r) = (D_{BS}(r)/\sigma)^{1/4}$ and
$T_{BH}(r) = (D_{BH}(r)/\sigma)^{1/4}$.\\
6) Use such temperature to integrate the luminosity (\ref{eq:luminosity})
$L_{BS}(\nu)$ and $L_{BH}(\nu)$ for several values of $\nu$.\\


\section{Boson Stars as mimickers of black holes}
\label{sec:spectra}

We use the simple model described above and look for the boson star configurations that are able to produce approximately the same power spectrum due to the presence of a black hole of the same mass. In order to do so, we first set two physical situations: 

\begin{itemize}
\item[A)] a black hole of 3$\times 10^9M_{\odot}$ with an accretion rate of $2\times 10^{-6}M_{\odot}/yr$.
\item[B)] a black hole of $10M_{\odot}$ with an accretion rate of $2\times 10^{-12}M_{\odot}/yr$.
\end{itemize}

\noindent The angle used in the calculations is $\vartheta=60$ degrees in all the cases presented, and the disk outer edge is assumed to be located at 50$M$. A different inclination of the disk would shift the power spectrum toward higher luminosities for small $\vartheta$ and smaller luminosities for $\vartheta$ approaching $\pi/2$ in all the frequencies.

\subsection{The unexpected case of $\Lambda=0$.}

In the past it was considered that BSs could be detected because the emission power spectrum from an accretion disk showed a hardening at high frequencies by orders of magnitude \cite{diego-acc}, however the boson star chosen was the most compact in the stable branch of the $\Lambda=0$ case; it was shown in \cite{Guzman2006} that by choosing a different boson star with a different compactness it was possible to mimic the spectrum. Here we show that the spectrum is mimicked independently of the physical scale at the price of changing the value of the boson mass.

In Fig. \ref{fig:spectra} we show the two physical situations A and B and mimic the black hole accretion disk spectra with one corresponding to a disk around an appropriate boson star configuration. For reference we also present the spectra calculated using the most compact configuration in the stable branch of the $\Lambda=0$ case (see Fig. 1) that corresponds to a boson star mass $M=0.633M_{pl}^{2}/m$. On top of the black hole spectrum there is the one due to a boson star configuration with $M=0.473M_{pl}^{2}/m$ on the same stable branch of boson star configurations. 
In the case A a mass of $3\times 10^{6}M_{\odot}$ and $M=0.473M_{pl}^{2}/m$ needs the mass of the boson to be $m=1.2\times 10^{-27}$GeV, whereas for the case B, a BHC mass of 10$M_{\odot}$ needs $m=2.51\times 10^{-22}$GeV.
That is, we show here that {\it it is actually possible to mimic a black hole at two very different scales at the price of changing the mass of the boson}. This indicates that restrictions on the value of $m$ should be in turn; at the moment, since no fundamental scalar field particle has ever been discovered the boson mass is a free parameter; once there are bounds on the mass of a boson particle the set of BHCs that can be mimicked with boson stars will be restricted, and the only free parameter will be the self-interaction.

\begin{figure}[htp]
\includegraphics[width=8cm]{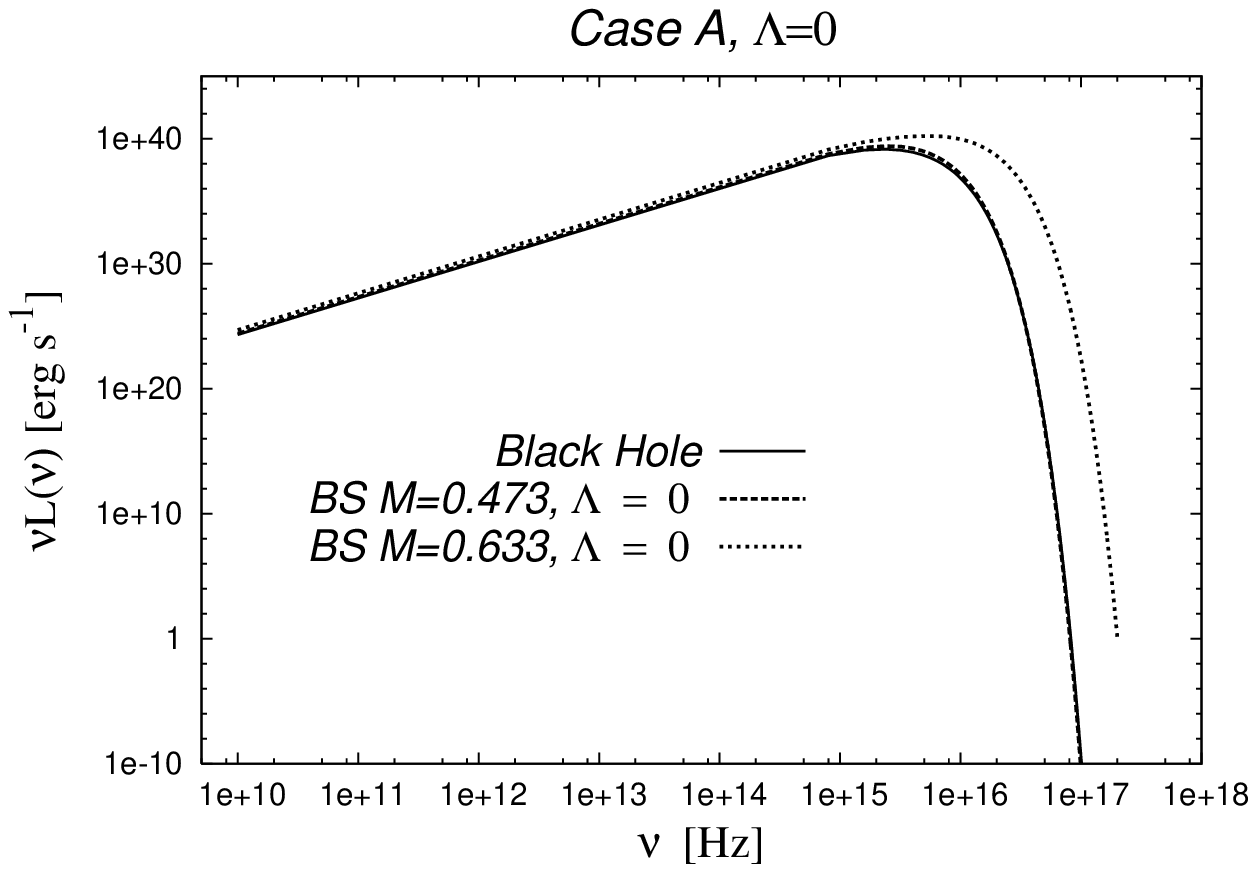}
\includegraphics[width=8cm]{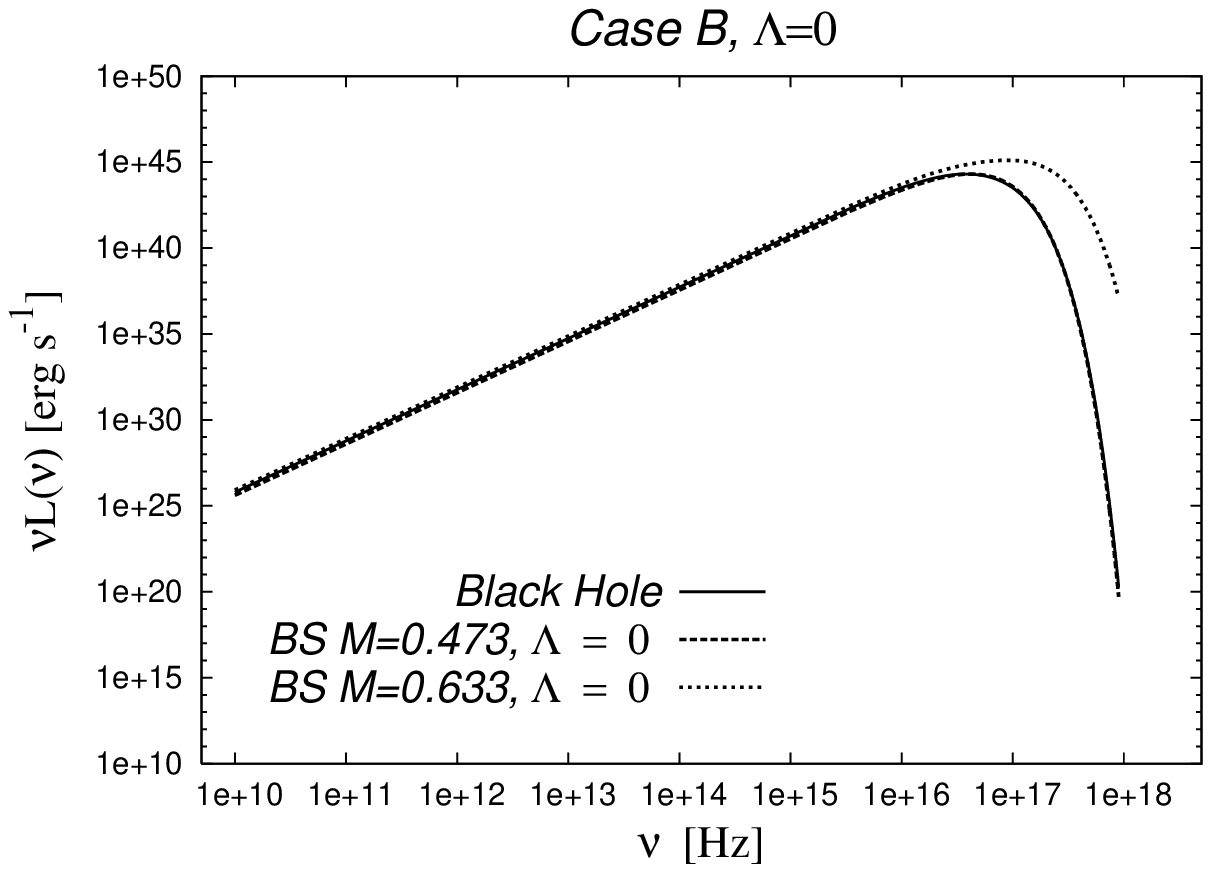}
\caption{\label{fig:spectra} The emission spectra for cases A (top) and B (bottom)  due to the black hole and two different boson stars: the most compact of the stable branch for $\Lambda=0$ case ($M=0.633M_{pl}^{2}/m$, see Fig 1.) and an adequate configuration ($M=0.473M_{pl}^{2}/m$) that presents a power spectrum that mimics that of the black hole. The most compact boson star is used only as a reference. The configuration with $M=0.473$ is used for the two different astrophysical situations. For case A: $m=1.2\times 10^{-27}$GeV and for case B: $m=2.51 \times 10^{-22}$GeV. The spectrum due to the black hole and to the mimicker lie approximatelyon top of each other. For similar physical configurations the spectrum is comparable to those obtained in \cite{Harko2009c}, where also the redshift was taken into account.}
\end{figure}

\subsection{The case $\Lambda \ne 0$.}

With the introduction of the self-interaction fourth order term in the potential, it happens that a new free parameter is involved, which makes easier to fit for instance the power spectrum of a disk, and is bad for the boson star model in terms of the number of parameters. However, if a boson is to be observed in the laboratory, the mass of it would be fixed and the $\Lambda$ parameter would become the only free parameter. Meanwhile here we present a few cases for $\Lambda \ne 0$ and indicate the range in which the boson stars as black hole mimickers should be searched in the configurations set of Fig. 1.

We consider again the physical situations A and B in order to show that the best mimicker is scale independent as in the case $\Lambda=0$, that is, we chose our best black hole mimicker from Fig. 1, for a given $\Lambda$ in one case and it will also work for other astrophysical cases with different masses of the central object and different accretion rates.

We choose the case $\Lambda=50$ and present the spectra as before for cases A and B. Again, there is a mimicker configuration with mass $M=0.9898M_{pl}^{2}/m$. In case A the mass of the boson  required for the boson star to be a mimicker is $m=1.88 \times 10^{-27}$Gev whereas for case B $m=5.25 \times 10^{-22}$GeV.

\begin{figure}[htp]
\includegraphics[width=8cm]{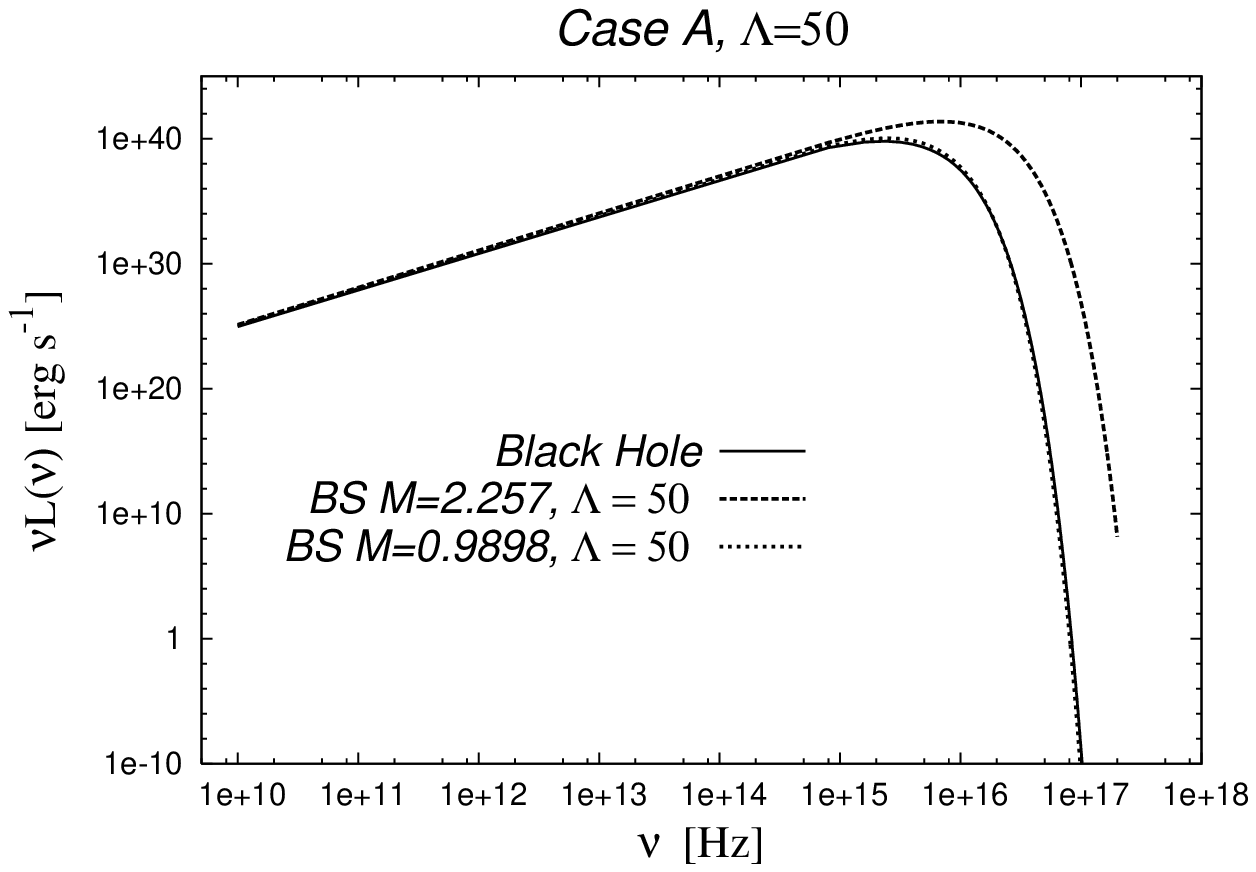}
\includegraphics[width=8cm]{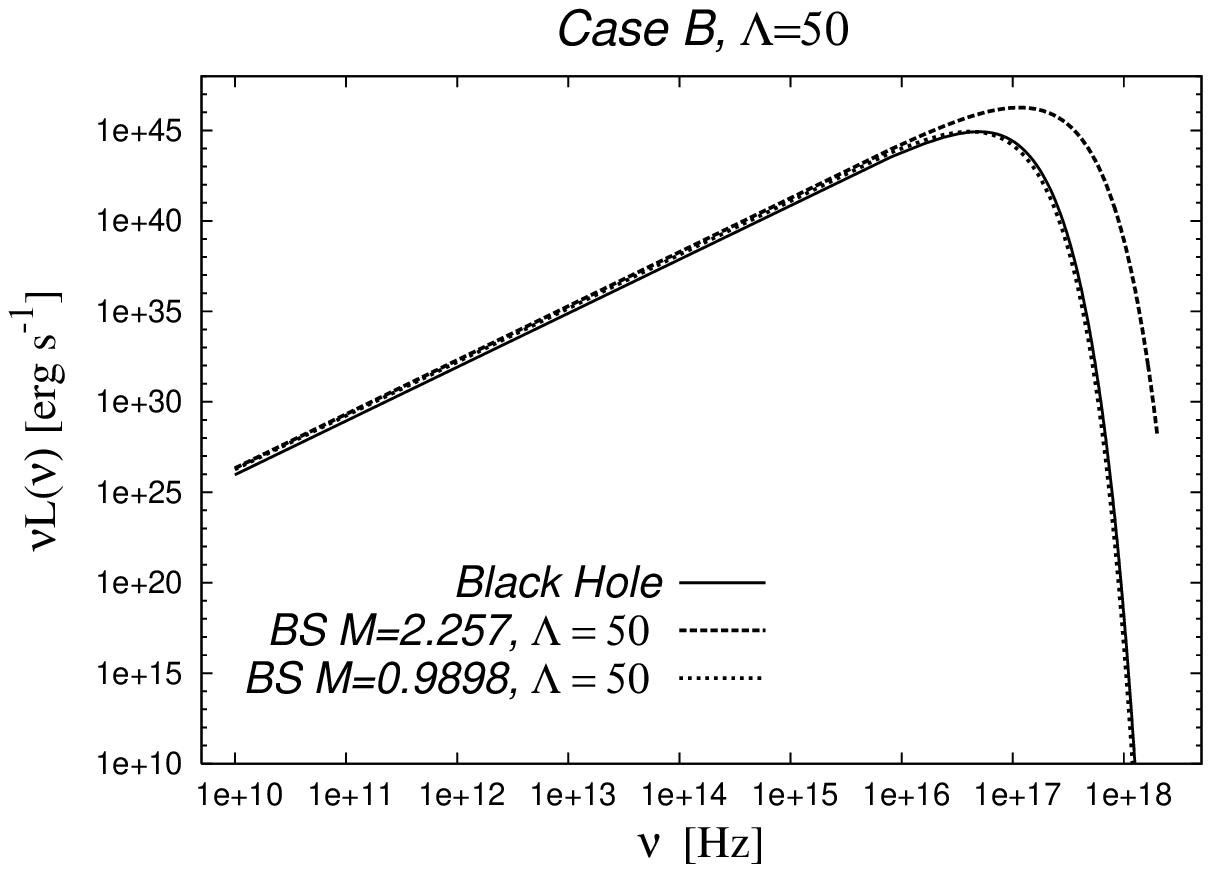}
\caption{\label{fig:spectra} The emission spectra for cases A (top) and B (bottom)  due to the black hole and two different boson stars: the most compact of the stable branch for $\Lambda=50$ case ($M=2.257M_{pl}^{2}/m$, see Fig 1.) and an adequate configuration ($M=0.9898M_{pl}^{2}/m$) that presents a power spectrum that mimics that of the black hole. As in the previous case, the same configuration with $M=0.9898$ is used for the two different astrophysical situations. For case A:  $m=1.88\times 10^{-27}$GeV and for case B: $m=5.25 \times 10^{-22}$GeV respectively. The spectrum due to the black hole and to the mimicker lie approximately on top of each other.}
\end{figure}

\subsection{Tracking all the cases.}

\begin{figure}[htp]
\includegraphics[width=8cm]{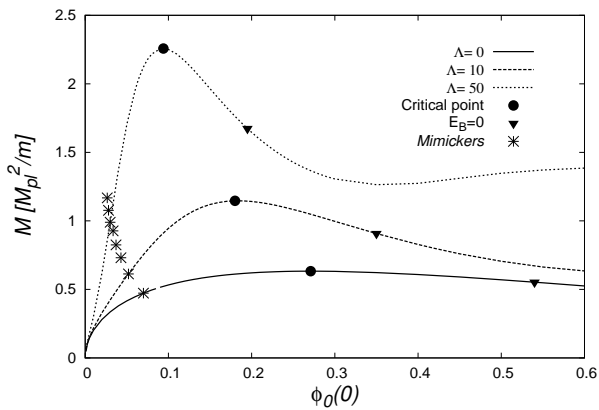}
\caption{\label{fig:mimickers} The mimicker configurations are indicated with stars for values of $\Lambda=0,10,20,30,40,50,60,70$. It is expected that for  every value of $\Lambda$ there is a mimicker configuration that lies near the ones indicated for the values shown here.}
\end{figure}

The mimickers based on this model of accretion disk have been located in the diagram of Fig. 1 for each value of $\Lambda$ as those configuration which spectrum approximates that of the black hole with the same mass, only in terms of the correct boson mass. In order to show where the mimickers we found are located among the whole set of boson star solutions, we present Fig. \ref{fig:mimickers} where we indicate the location of the mimickers and their apparently continuous distribution.

It is natural to expect that the mimicker should change if the disk model is different, if the angle of inclination used to calculate the luminosity is different and if other effects are taken into account. Nevertheless, our results show that the boson star mimickers presented here are a set of measure zero among the total set of stable boson stars with different values of $m$ and $\Lambda$ out of which correct mimickers can be chosen.

\begin{figure}[htp]
\includegraphics[width=8cm]{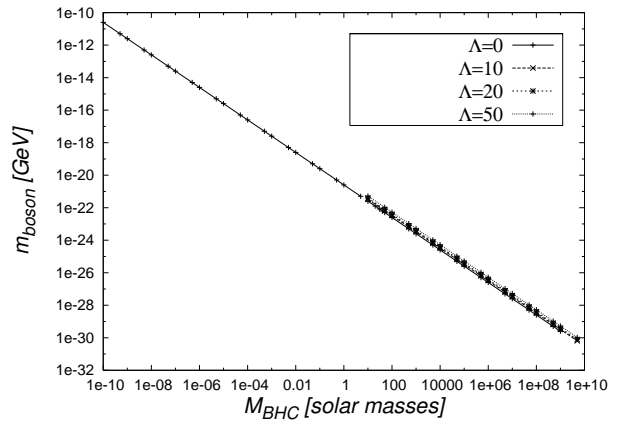}
\caption{\label{fig:boson_masses} We show the mass of the boson in GeVs in terms of the mass of the black hole candidate in solar masses for mimicker configurations and various values of $\Lambda$. We extend the case $\Lambda=0$ case in order to show that for the axion mass case the object would be very small.}
\end{figure}

Another point has to do with the boson masses in terms of the mass of the BHC. We show in Fig. \ref{fig:boson_masses} the boson mass scaling with the mass of the black hole candidate for mimicker configurations. As an example, assuming the axion mass is of the order of 1meV ($10^{-12}$GeV) there is no astrophysical configuration even about 1 $M_{\odot}$ that might possibly act as an astrophysical black hole mimicker; instead the mass of the object would be $10^{-8}M_{\odot}$.
Another example is the Higgs boson, with a mass bounded by $m_{Higgs}>115$GeV would imply black hole mimickers with mass $M \sim 10^{-22}M_{\odot}$.

\section{Lensing, a way to distinguish a black hole from its mimicker}
\label{sec:lensing}

In order to provide a method to distinguish a boson star mimicker from a black hole we show the lensing deflection effects for a boson star mimicker configuration and the black hole mimicked. For this we choose to solve the full geodesics equation of null rays for the mimicker and the black hole

\begin{equation}
\frac{d^2x^{i}}{d \gamma ^2} +\Gamma^{i}{}_{jk}\frac{dx^{j}}{d\gamma} \frac{dx^{k}}{d\gamma}=0,
\label{eq:geos}
\end{equation}

\noindent where $x^i=(t,r,\varphi)$ are respectively the time, radial, azimutal angular coordinate, and $\gamma$ is the affine parameter of the geodesics. We integrate the geodesic equations numerically using a fourth order accurate Runge-Kutta integrator in terms of the affine parameter. For the black hole space-time we use de Schwarzschild solution in Schwarzschild coordinates. For the boson star, since the metric functions are calculated as a numerical solution of equations (\ref{sphericalekgc-sc-rescaled}), we solve the geodesic equation by sourcing the ODE integrator with second order interpolated values of the Christoffel symbols in (\ref{eq:geos}). In the calculations of the null ray trajectories we have verified the convergence of the numerical method by monitoring the null condition of the tangent vector of the geodesic at every point of the space.

\begin{figure}[htp]
\includegraphics[width=8cm]{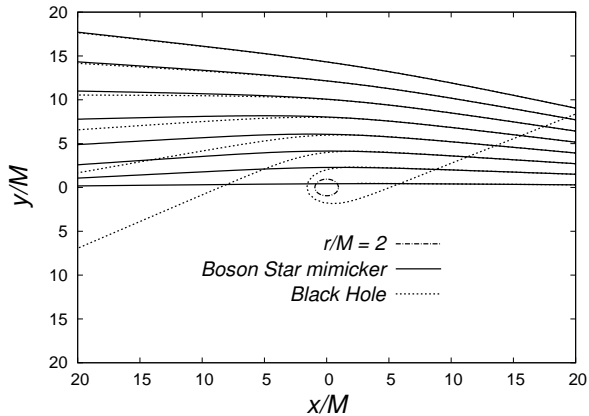}
\caption{\label{fig:deflection} We show a set of null geodesics for the black hole and its boson star mimicker. We choose the mimicker for $\Lambda=0$. We use units for the coordinates on the equatorial plane $x$ and $y$ scaled with $M$ (the mass of the BHC) and therefore this plot corresponds to both cases A and B for $\Lambda=0$, where the scales for cases A and B are obtained when the value of $M$ of the BHC is substituted.}
\end{figure}

\begin{figure}[htp]
\includegraphics[width=8cm]{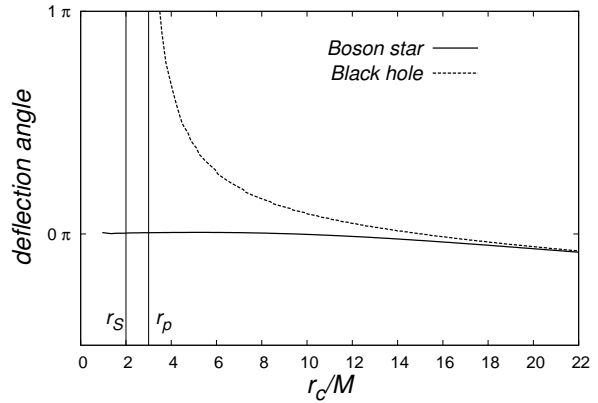}
\caption{\label{fig:photon_sphere} We show the deflection angle of the null rays for the black hole and its boson star mimicker corresponding again to the case A and $\Lambda=0$ presented above. In this plot $r_p$ is the photon sphere radius of the black hole, $r_{S}$ is the Schwarzschild radius, and the label $r_c$ indicates the closest approach radius. We only show the deflection angle for $r>M$ because in spherical coordinates the geodesic equations become stiff; a detailed study of this region would need a patch with different coordinates which is not within the scope of this paper.}
\end{figure}

We use various values of the impact parameter and calculate the deflection angle in both space-times (the black hole and the boson star). The results are shown in Fig. \ref{fig:deflection} as a set of null rays on the equatorial plane for the black hole and its mimicker for the case $\Lambda=0$. It can be seen that for small values of the impact parameter the black hole traps the null rays, whereas the boson star is transparent and only deflects the null rays.
The deflection angle in terms of the closest approach to the center is shown in Fig. \ref{fig:photon_sphere}, where it is shown the expected result that boson stars do not have a photon sphere since they have no horizon, that is, there is no minimum impact parameter after which the null ray trajectories are trapped.

In the example of Fig. \ref{fig:deflection} the luminous source is located at a distance of $50M$. This distance can be extended arbitrarily far away from the BHC. From Fig. 7 it can be learned that if the deflection angle can be measured with observational resolution of $r \sim 15-20M$, a boson star mimicker can be distinguished from a black hole because the curves are clearly distinct. On the other hand, because the deflection angle due to the boson star mimicker starts being on top of that of the black hole for $r > 20M$, these objects cannot be distinguished from each other using the deflection angle in this regime.

\section{Conclusions}
\label{sec:conclusions}

We have shown that stable spherically symmetric boson stars are black hole mimickers when the emission spectrum of a simple accretion disk model around a black hole is compared to another around a boson star of the same mass. 

For each value of the self-interaction coefficient (including the free field case ($\Lambda=0$), given astrophysical parameters for the mass of the black hole candidate and the accretion rate parameter, it was possible to find a stable boson star configuration whose accretion disk shows the equivalent spectrum for a disk around a black hole of the same mass.

We have also shown that for each value of the self-interaction coefficient $\Lambda$, once the mimicker boson star has been found for a given astrophysical combination with mass of the BHC and accretion rate, that same boson star configuration is the mimicker for other astrophysical parameters combination.

Most importantly, we also showed the way a boson star could be distinguished from a black hole by studying the light deflection. Considering that the black hole mimicker property of boson stars can be extended to other accretion disk models, it is possible that the light deflection could be together with the gravitational wave emission \cite{Lehner2007,ruxandra}, one of the few experiments that could discard or confirm the existence of boson stars.


\section*{Acknowledgments}

This research is partly supported by 
grants: 
research program FC-UAMeX,
CIC-UMSNH 4.9, 
PROMEP UMSNH-CA-22 
and CONACyT 79995. 



\end{document}